\documentclass[aps,prb,twocolumn,groupedaddress,10pt]{revtex4-1}
\usepackage{graphicx}
\usepackage{color}
\begin{document}

\title{Interlayer excitonic insulator in two-dimensional double-layer semiconductor junctions:
An explicitly solvable model}

\author{Maxim Trushin}
\affiliation{Institute for Functional Intelligent Materials, National University of Singapore, Singapore 117544}
\affiliation{Department of Materials Science and Engineering, National University of Singapore, Singapore 117575}
\affiliation{Centre for Advanced 2D Materials, National University of Singapore, Singapore 117546}

\date{\today}

\begin{abstract}
Excitonic insulators conduct neither electrons nor holes but bound electron-hole pairs, excitons.
Unfortunately, it is not possible to inject and detect the electron and hole currents independently within a single semiconducting layer.
However, {\em interlayer} excitonic insulators provide a spatial separation of electrons and holes enabling exciton current measurements.
The problem is that the spatial separation weakens electron-hole pairing and may lead to interlayer exciton disassociation.
Here we develop an explicitly solvable model to determine an interlayer separation that
is strong enough to prevent electron and hole hopping across the layers but still allows for electron-hole pairing
sufficient for transition into an interlayer excitonic insulator state.
An ideal junction to realize such a state would comprise a pair of identical narrow-gap two-dimensional semiconductors separated 
by a wide-gap dielectric layer with low dielectric permittivity. 
The present study quantifies parameters of such a junction by taking into account interlayer coherence effects.
\end{abstract}

\maketitle

\section{Introduction}
\label{intro}

The concept of excitonic insulator (EI) dates back to the 60's when
the normal insulating ground state was found to be unstable against the formation of electron-hole
bound states (excitons) in semiconductors with a narrow bandgap \cite{keldysh1965possible,CLOIZEAUX1965259,jerome1967excitonic}. 
The instability emerges as soon as the exciton binding energy exceeds the semiconducting bandgap.
The resulting state remains insulating for holes and electrons separately but becomes capable to conduct excitons.
 Unfortunately, low exciton binding energy and lack of separate control over electron and hole populations 
conceal manifestations of the EI state in bulk semiconductors
\cite{PRL1991bucher,PRL2007cercellier,PRL2009wakisaka,lu2017zero,PRL2017gapcontrol}.
However, the recent advent of two-dimensional (2D) materials has revived the field and led to the interlayer excitonic insulator (IEI) concept
\cite{PRB2019InAs,PRB2021QMC,PRB2012RPA,Brunetti_2018,shi2022bilayer,du2017evidence,ma2021strongly,zhang2022correlated}.

\begin{figure}
 \includegraphics[width=\columnwidth]{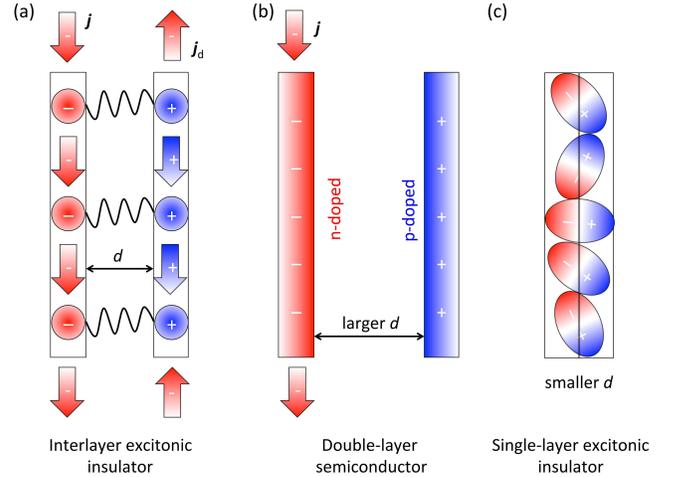}
 \caption{(a) Interlayer excitonic insulator in a drag-counterflow geometry \cite{NatPhys2008macdonald} with the current densities $\mathbf{j}$ and $\mathbf{j}_d$ shown.
 The interlayer spacer thickness, $d$, must be chosen within a certain value range. (b) If $d$ is too large, then electrons and holes are not paired.
 (c) If $d$ is too small, then the spacer cannot prevent charge hopping across the layers disabling the drag-counterflow measurements.
} 
 \label{fig1}
\end{figure}

The idea is to make use of a double-layer semiconductor structure with a dielectric spacer that prevents the interlayer electron-hole pairs from recombination but allows for strong Coulomb pairing \cite{ma2021strongly,zhang2022correlated,PRB2019InAs,PRB2021QMC,PRB2012RPA,macdonald-interlayer-tunnel,Brunetti_2018}.
Besides higher exciton binding energies in 2D semiconductors, the double-layer configuration makes it possible to realize a drag-counterflow setup \cite{NatPhys2008macdonald,eisenstein2004bose,vignale-macdonald-1996}
with two pairs of contacts for independent control of electron and hole transport, see Fig. \ref{fig1}(a). 
However, the requirements for a suitable dielectric spacer are somewhat contradictory.
On the one hand, the electron-hole attraction across the junction must be much weaker than the intralayer confinement to avoid interlayer charge hopping.
On the other hand, the electron-hole interactions must be sufficiently strong to ensure stability of the IEI phase state.
Indeed, to achieve better interlayer electrical isolation, one could increase the spacer thickness $d$, see Fig. \ref{fig1}(b).
However, increasing $d$ leads to strong reduction of the bare Coulomb 2D Fourier transform, $V_q=2\pi/q$,
by the form-factor $\mathrm{e}^{-qd}/\epsilon$, where $\epsilon$ is the dielectric permittivity of the interlayer media.
The reduction is especially strong for larger in-plane wave vectors $\mathbf{q}$ relevant for tightly bound excitons \cite{TightlyEx2019}.
To achieve stronger electron-hole pairing one could decrease $d$, see Fig. \ref{fig1}(c).
This increases the interlayer charge hoping probability and gradually reduces the double-layer structure to a bilayer material hosting conventional excitons \cite{jia2022evidence}.
Hence, even if the IEI state exists at all, it remains stable only within a certain interval of values $d$ limited from below and above by material parameters. 

The quantum mechanical effects add even more interesting physics into the IEI problem.
The eigenstate of a charge carrier in a double-layer structure obviously does not coincide with that of a separated layer. 
Once an electron (or a hole) is created in an eigenstate of a given layer its further evolution is governed by the double-layer Hamiltonian.
The resulting probability density oscillates between two layers, and
at certain time points its maximum occurs on the opposite side
of a {\em symmetric} double-layer structure.
Hence, the electrons and holes injected into the eigenstates of the separated layers can hop between the layers when evolving in time.
The phenomenon could be seen as an interlayer coherence
that can be suppressed by either the double-layer asymmetry or disorder.
%As interlayer excitons are supposed to be made of electrons and holes localised in their respective layers, they cannot exist 
%in a perfectly symmetric double-layer structure without disorder limiting the interlayer coherence effects.

The main question addressed in the present paper is whether the interlayer coherence between electrons and holes 
is beneficial for bringing them into the IEI state suitable for the drag-counterflow measurements, Fig. \ref{fig1}(a).
To answer this question we maximize the coherence effect
by employing a perfectly symmetric double-layer structure modeled by a double-delta-shaped out-of-plane confinement. 
We reveal two competing mechanisms: (i) electron-hole pairing with larger in-plane wave vectors 
that facilitates transition into the IEI state, and
(ii) interlayer hopping that hampers formation of the IEI state.
We find the set of parameters at which the mechanism (i) dominates in symmetric double-layer structures
and makes transition into the IEI state possible.

The drag-counterflow setup implies no superfluidity that is in line with the Kohn-Sherrington classification \cite{Kohn1970}
relating the electron-hole bound complexes to type II bosons.
The possibility of exciton condensation and superfluidity in electron-hole double layers has been discussed in 
Refs. \cite{Littlewood1995,vignale-macdonald-1998}.

The paper is organized as follows. Section \ref{model} introduces the one-particle framework
for understanding physics of a double-layer semiconductor junction.
Section \ref{manybody} builds up with a mean-field theory of electron-hole pairing to describe
transition between the normal and IEI states.
Section \ref{discuss} provides discussion of asymmetric double-layer structures with 
different relative permittivities of the dielectric spacer. Section \ref{outro}
concludes with a recipe for the IEI using existing 2D materials.

\section{Single-particle prerequisites}
\label{model}

\begin{figure}
 \includegraphics[width=\columnwidth]{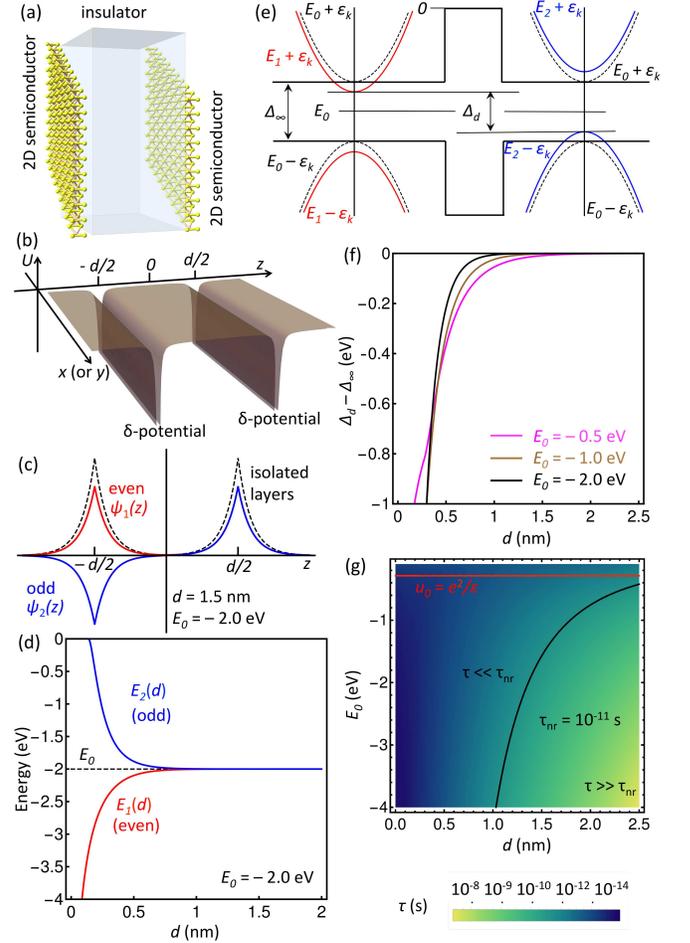}
 \caption{Non-interacting quantum mechanics of a double-layer semiconductor junction.
 (a) A pair of identical 2D semiconducting layers separated by an insulator.
 (b) Double-delta-shaped out-of-plane potential as a model for the double-layer junction.  
 (c) Even and odd states of the the double-layer junction.
 (d) As the layers are getting closer to each other,
 the even states become deeper shifting the 2D bands down in energy, whereas the effect on the odd states is opposite.
(e) Band shifting in the semiconductor layers due to their proximity to each other.
The energy is counted from the bottom of the conduction band in the dielectric spacer.
(f) Interlayer bandgap reduction upon interlayer proximity for different depths of the potential well $E_0$.
(g) Interlayer coherence time estimated by Eq. (\ref{tau})
for different energy depths, $E_0=-m_0u_0^2/(2\hbar^2)$, and interlayer distances, $d$.
The depth must obviously be lower than $-m_0 e^4/(2\epsilon^2\hbar^2)$ 
with $\epsilon=6.9$ for $h$-BN (red line),
and $\tau$ must be substantially larger than exciton life-time 
(typical defect-assisted non-radiative recombination time, black curve)
to keep electrons and holes in their respective layers.
} 
 \label{fig2}
\end{figure}

Let us first describe the double-layer junction at a single-particle level.
The junction comprises two identical 2D semiconductors separated by a dielectric layer of thickness $d$, see Fig. \ref{fig2}(a,b).
Each semiconducting layer is described by the effective low-energy 2D Hamiltonian 
resulting in the dispersion of a massive Dirac particle \cite{Andor2015,PRL2012xiao}.
As the semiconductors are 2D, the out-of-plane confinement must be very narrow for each layer.
Such ultimately narrow confinements are conveniently described by the Dirac delta-shaped potentials.
%The junction must be described as a whole, rather than as a stack of weakly coupled layers, to take into account quantum mechanical effects properly.
Thus, the model Hamiltonian is written as $\hat H= \hat H_\perp + \hat H_\parallel$, where
\begin{equation}
 \hat H_\parallel=\left(
\begin{array}{cc}
\Delta_\infty/2 & \hbar v (\hat k_x-i \hat k_y)\\
\hbar v (\hat k_x+i\hat k_y) &  -\Delta_\infty/2
   \end{array}
   \right),
   \label{Hparallel}
\end{equation}
\begin{equation}
 \hat H_\perp =\left(
\begin{array}{cc}
\frac{\hbar^2 \hat k_z^2}{2m_0}+U(z)& 0\\
0 & \frac{\hbar^2 \hat k_z^2}{2m_0}+U(z) 
   \end{array}
   \right),
   \label{Hperp}
\end{equation}
 with $U(z)$ given by
\begin{equation}
 U(z)= - u_0\left[ \delta\left(z-\frac{d}{2}\right)
 +  \delta\left(z+\frac{d}{2}\right)\right].
 \label{U}
\end{equation}
Here, $\hbar \hat k_{x,y,z}$ are the components of the electron momentum operator $\hbar \hat \mathbf{k}$
with $\hbar$ being the Planck constant,
$m_0$ is the free electron mass, $v$ is the band parameter (effective velocity),
$u_0$ is the layer confinement parameter, and $\Delta_\infty$ is the bandgap at $d\to\infty$.

The in-plane term, $\hat H_\parallel$, can be deduced from the minimal k.p model describing the coupled dynamics of the 
valence and conduction bands in 2D transition metal dichalcogenides \cite{Andor2015,PRL2012xiao}. 
Written in the sublattice basis, the low-energy expansion of the k.p Hamiltonian results in the off-diagonal terms linear in $\hat k_{x,y}$.
The spin-orbit splitting, electron--hole asymmetry, and the trigonal warping are neglected here.
The out-of-plane electron motion is described by $\hat H_\perp$ in terms of free electron mass
because there is no periodicity along $z$-axis and no effective electron mass can be introduced.
However, the out-of-plane motion is restricted by the layer confinement, hence, $k_z$ is not a good quantum number. 

The eigenfunctions of $\hat H$ can be factorized and written explicitly as 
$\Psi_{1,2}^\pm=\psi_{1,2}(z)\chi^\pm(x,y)$. Here,
the indices ``1,2'' stand respectively for 
the even and odd states, see Fig. \ref{fig2}(c,d), and
``$\pm$'' refers to the conduction and valence bands, see Fig. \ref{fig2}(e).
The even/odd classification refers only to the symmetric double-layer structures
considered in sections \ref{model} and \ref{manybody}. 
If the symmetry was broken, then the indices ``1,2'' would respectively refer to the left/right 
layer of the junction, see section \ref{discuss}.
The factorized functions read \cite{santarsiero2019flea,ahmed2016revisiting}
\begin{equation}
 \psi_{1,2}(z)=\frac{B_{1,2}}{\sqrt{2}}\left\{
\begin{array}{ll}
(1\pm \mathrm{e}^{\kappa_{1,2}d})\mathrm{e}^{\kappa_{1,2}z}, & z\leq -\frac{d}{2};\\
\mathrm{e}^{\kappa_{1,2}z} \pm \mathrm{e}^{-\kappa_{1,2}z}, & -\frac{d}{2}< z <\frac{d}{2};\\
\pm(1\pm \mathrm{e}^{\kappa_{1,2}d})\mathrm{e}^{-\kappa_{1,2}z}, & z\geq \frac{d}{2};
\end{array}
\right.
\end{equation}
where $B_{1,2}=\sqrt{\kappa_{1,2}/(\mathrm{e}^{\kappa_{1,2}d}\pm\kappa_{1,2}d\pm 1)}$,
\begin{equation}
\kappa_{1,2} = \frac{m_0u_0}{\hbar^2}+\frac{1}{d}W_0\left(\pm\frac{m_0u_0d}{\hbar^2}
\mathrm{e}^{-\frac{m_0u_0d}{\hbar^2}}\right),
\end{equation}
with $W_0$ being the Lambert function (ProductLog in Wolfram's Mathematica), and
\begin{equation}
\chi^+(x,y)=\frac{1}{L}\mathrm{e}^{ik_x x+ik_y y}
\left(
\begin{array}{c}
\cos\frac{\gamma}{2}\\
\sin\frac{\gamma}{2}\mathrm{e}^{i\phi}
\end{array}\right),
\label{chi+}
\end{equation}
\begin{equation}
\chi^-(x,y)=\frac{1}{L}\mathrm{e}^{ik_x x+ik_y y}
\left(
\begin{array}{c}
\sin\frac{\gamma}{2}\\
-\cos\frac{\gamma}{2}\mathrm{e}^{i\phi}
\end{array}\right),
\label{chi-}
\end{equation}
where $\tan\gamma=2\hbar v k/\Delta_\infty$, $\tan\phi=k_y/k_x$, $k=\sqrt{k_x^2+k_y^2}$, and $L$ is the layer size.
The corresponding eigenvalues are given by
$E_{1,2}^\pm=E_{1,2}(d)\pm\varepsilon_k$, where
\begin{equation}
 \varepsilon_k=\sqrt{(\hbar v k)^2 +(\Delta_\infty/2)^2},
\end{equation}
and
\begin{equation}
 E_{1,2}(d) = E_0 \left[1+\frac{\hbar^2}{m_0 u_0 d}W_0\left(\pm\frac{m_0u_0d}{\hbar^2}
\mathrm{e}^{-\frac{m_0u_0d}{\hbar^2}}\right) \right], 
\end{equation}
with $E_0=-m_0u_0^2/(2\hbar^2)$. The two energy branches, $E_{1,2}(d)$, merge to $E_0$ in the limit of $d\to\infty$, see Fig. \ref{fig2}(d). 
As $d$ decreases, the bands shift in opposite directions forming
a type II junction that potentially can host interlayer excitons \cite{ozcelik2016band}.
Note, however, that we consider a homojunction, not a heterostructure \cite{ozcelik2016band}.
The interlayer bandgap reads $\Delta_d = \Delta_\infty - E_2(d)+E_1(d)$, see Fig. \ref{fig2}(f).
It naturally reduces when the layers get closer to each other.
If the Fermi level is fixed, then the left semiconducting layer becomes n-doped, whereas the right one
acquires p-doping. Note, that such a doping-by-proximity effect is intrinsic for our model.

The single-particle model is able to indicate the mechanisms that can potentially hamper the IEI formation.
First of all, the electrons and holes should sit deeply in the respective layers
to prevent recombination caused by their mutual attraction.
Neglecting dependence on $d$, we can estimate the critical $u_0$ as
$\sim e^2/\epsilon$ that results in the desirable depth $E_0 \ll -m_0 e^4/(2\epsilon^2 \hbar^2)$.
This is not a strong criterion in the presence of a dielectric spacer with $\epsilon\gg 1$,
see the red line in Fig. \ref{fig2}(g).

It is instructive to consider the quantum mechanical effects leading to 
electron-hole interlayer hopping.
The effect of quantum mechanical superposition is especially
obvious when the two layers are perfectly identical.
In this case, an electron (or a hole) is not localized in either layer.
The position probability density $|\Psi_{1,2}^\pm|^2$ is symmetric 
with respect to $z=0$ for both states 1 and 2,
meaning that a position measurement would reveal an electron (or a hole) with the same probability in either layer. 

An electron (or a hole) state created at time $t=0$ in a given layer
involves a superposition between $\psi_1(z)$ and $\psi_2(z)$.
To be specific, consider an electron localized in the left layer
and a hole localized in the right layer described, respectively, by the 
wave functions
\begin{eqnarray}
 \Psi_L^e(x,y,z) &= &\frac{1}{\sqrt{2}}\left[\psi_1(z)-\psi_2(z)\right]\chi^+(x,y),\\
 \Psi_R^h(x,y,z) &= &\frac{1}{\sqrt{2}}\left[\psi_1(z)+\psi_2(z)\right]\chi^-(x,y),
\end{eqnarray}
see Fig. \ref{fig2}(c) for $\psi_{1,2}(z)$ profiles.
The states are not stationary, and they evolve in accordance with the standard solutions
of the time-dependent Schr\"odinger equation written as
\begin{eqnarray}
  && \Psi_L^e(x,y,z,t) = \\
\nonumber  && \frac{1}{\sqrt{2}}\left[\psi_1(z){\mathrm e}^{-iE_1^+t/\hbar}
 -\psi_2(z) {\mathrm e}^{-iE_2^+t/\hbar}\right]\chi^+(x,y),\\
 && \Psi_R^h(x,y,z,t) = \\
  \nonumber && \frac{1}{\sqrt{2}}\left[\psi_1(z){\mathrm e}^{-iE_1^-t/\hbar}
 +\psi_2(z){\mathrm e}^{-iE_2^-t/\hbar}\right]\chi^-(x,y).
\end{eqnarray}
Obviously, the probability density $|\Psi_{L,R}^{e,h}(x,y,z,t)|^2$
oscillates with the period given by
\begin{equation}
\tau(d)=\frac{2\pi \hbar}{E_2(d)-E_1(d)}.
\label{tau}
\end{equation}
The interlayer probability density oscillation period could also 
be seen as an interlayer coherence time.
Within the period $\tau(d)$, the electron (or hole) probability density maximum hops back and forth
between the layers. Hence, an electron (or a hole) injected into one of the two layers of
a symmetric double-layer structure can be found 
in any layer at a random time point $t\gg \tau(d)$. 

Microscopically, the quantum mechanical ``measurement''
takes place each time when a charge carrier is trapped by a defect.
The process is usually associated with non-radiative exciton recombination 
and occurs in 2D semiconductors \cite{MoS2-2015,WS2-2015} at the time scale
$\tau_\mathrm{nr} \sim$  $10^{-12}$ -- $10^{-10}$ s. 
Obviously, if $\tau_\mathrm{nr}\gg \tau(d)$, then electrons and holes
have already hoped between the layers many times before recombining.
This is the case when the interlayer spacer is thin, see Fig. \ref{fig2}(g).
If EI state can form at all in such conditions, then
it should be seen as a single-layer EI, where
drag-counterflow measurements are impossible, see Fig. \ref{fig1}(c).
In contrast, if $\tau_\mathrm{nr}\ll \tau(d)$, then
electrons and holes remain in the respective layers within the excitonic lifetime,
and an EI state, if formed, could be detected in drag-counterflow measurements.

\section{Many-body model}
\label{manybody}

\begin{figure}
 \includegraphics[width=\columnwidth]{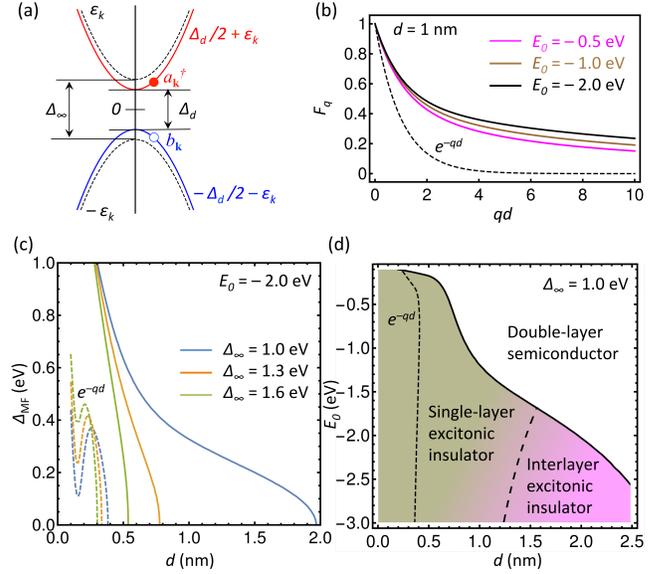}
 \caption{Mean-field theory of an interlayer excitonic insulator state emerging
 in a symmetric double-layer semiconductor junction.
 (a) Two-band reformulation of the four-band problem shown in Fig. \ref{fig1}(e),
 with the zero energy level placed in the midgap. An interlayer electron-hole pair is shown in terms
 of the respective creation operators, see Eq. (\ref{order-delta}).
 (b) The wave function overlap facilitates the Coulomb interactions
 between electrons and holes, as compared with the conventional case, $F_q=\mathrm{e}^{-qd}$,
 when electrons and holes are localized in the respective layers. The curves are given by Eq. (\ref{Fq}).
 (c) Mean-field many-body gap vs. interlayer distance for 
 different semiconductor gap values. The dashed curves show solutions in
 the conventional case with $F_q=\mathrm{e}^{-qd}$.
 (d) Phase diagram in terms of the interlayer distance and potential depth.
 The black long-dashed line separating single-layer and interlayer EI regions 
 is taken form Fig. \ref{fig2}(g). There is no sharp transition between the single-layer and interlayer EI states,
 as indicated by the gradient fill.
} 
 \label{fig3}
\end{figure}

We are now ready to write a mean-field Hamiltonian describing the many-body IEI state.
To do that, we reduce the four-band model shown in Fig. \ref{fig2}(e)
to a two-band one shown in Fig. \ref{fig3}(a) because $\Delta_\infty$ is supposed to be always 
large enough to prevent formation of the intralayer EI.
The effective model involves one conduction and one valence band
hosting electrons and holes in the single-particle states $\Psi_1^+$ and $\Psi_2^-$, respectively.
We symmetrize the bands placing the zero-energy level in the middle of the interlayer bandgap.
The mean-field Hamiltonian can be then written as \cite{Littlewood2004}
\begin{equation}
 \label{MF}
 H_\mathrm{MF}=\sum\limits_\mathbf{k}
 \left(a_\mathbf{k}^\dagger b_\mathbf{k}^\dagger\right)
 \left(
\begin{array}{cc}
\xi_\mathbf{k}  &   -\Delta_\mathbf{k}^\dagger\\
-\Delta_\mathbf{k} & -\xi_\mathbf{k}
\end{array}
\right)
\left( 
\begin{array}{c}
a_\mathbf{k} \\
b_\mathbf{k}
\end{array}
\right),
\end{equation}
where $a_\mathbf{k}^\dagger$ ($b_\mathbf{k}^\dagger$) are the electron creation operators
in the conduction (valence) band,
$a_\mathbf{k}$ ($b_\mathbf{k}$) are the respective hole creation operators,
$\xi_\mathbf{k}=\varepsilon_k+(E_1(d)-E_2(d))/2$,
and the mean-field parameter reads \cite{Littlewood2004}
\begin{equation}
 \Delta_\mathbf{k}  =  \sum\limits_{\mathbf{k}'}
 |V_{\mathbf{k}\mathbf{k}'}|\langle a_{\mathbf{k}'}^\dagger b_{\mathbf{k}'} \rangle.
 \label{order-delta}
 \end{equation} 
Here, $V_{\mathbf{k}\mathbf{k}'}$ is the electron-hole interaction matrix element.
Using the Bogolubov transformation
\begin{eqnarray}
a_\mathbf{k} & = & c_{1\mathbf{k}} \cos\frac{\zeta_\mathbf{k}}{2}  - c_{2\mathbf{k}} \sin\frac{\zeta_\mathbf{k}}{2}, \\
b_\mathbf{k} & = & -c_{1\mathbf{k}} \sin\frac{\zeta_\mathbf{k}}{2}  - c_{2\mathbf{k}} \cos\frac{\zeta_\mathbf{k}}{2},
\end{eqnarray}
with $\tan\zeta_\mathbf{k} =\Delta_\mathbf{k}/\zeta_\mathbf{k}$,
we arrive at the canonical form of the mean-field Hamiltonian given by
\begin{equation}
 \label{MF-2}
 H_\mathrm{MF}=\sum\limits_\mathbf{k}
 \sqrt{\xi_\mathbf{k}^2 + \Delta_\mathbf{k}^2}
 \left( c_{1\mathbf{k}}^\dagger c_{1\mathbf{k}} - c_{2\mathbf{k}}^\dagger c_{2\mathbf{k}}\right).
\end{equation}
In the low-temperature limit, the mean-field order parameter reads
\begin{equation}
 \Delta_\mathbf{k} = \frac{1}{2}\sum\limits_{\mathbf{k}'} |V_{\mathbf{k}\mathbf{k}'}|
 \frac{\Delta_{\mathbf{k}'}}{\sqrt{\xi_{\mathbf{k}'}^2 + \Delta_{\mathbf{k}'}^2}}.
 \label{MF-3}
\end{equation}
Equation (\ref{MF-3}) is formally equivalent to the gap equation derived in the seminal paper \cite{jerome1967excitonic}.
However, $V_{\mathbf{k}\mathbf{k}'}$ and $\xi_{\mathbf{k}}$ both depend on $d$ in our case.
Finally, we assume that the order parameter does not depend
on $\mathbf{k}$ and represents the mean-field bandgap, $\Delta_\mathrm{MF}$, which can be found from 
the gap equation written as
\begin{equation}
  \frac{1}{2}\sum\limits_{\mathbf{q}}
 \frac{V_q}{\sqrt{\xi_{\mathbf{q}}^2 + \Delta_\mathrm{MF}^2}}=1.
 \label{MF-4}
\end{equation}
Evaluation of $V_q$ must take into account
the wave function overlap between the single particle states
$\Psi_1^+$ and $\Psi_2^-$.
The two-particle wave function can be written as an antisymmetric combination of
the single-particle states given by
\begin{eqnarray}
 \nonumber \Psi({\mathbf{k}_1,\mathbf{r}_1;\mathbf{k}_2,\mathbf{r}_2}) &= & \frac{1}{\sqrt{2}}
 \left[\Psi_1^+(\mathbf{k}_1,\mathbf{r}_1)\Psi_2^-(\mathbf{k}_2,\mathbf{r}_2) \right. \\
 && \left. -\Psi_1^+(\mathbf{k}_2,\mathbf{r}_2)\Psi_2^-(\mathbf{k}_1,\mathbf{r}_1) \right],
\end{eqnarray}
where $\mathbf{r}_{1,2}=(x_{1,2},y_{1,2},z_{1,2})$ are the coordinates of particles 1 and 2,
and $\mathbf{k}_{1,2}$ are their in-plane wave vectors.
Transition from the state with $\mathbf{k}_1$, $\mathbf{k}_2$ to the state
with $\mathbf{p}_1$, $\mathbf{p}_2$ is described by the following matrix element
\begin{eqnarray}
\nonumber V_{\mathbf{p}_1\mathbf{p}_2\mathbf{k}_1\mathbf{k}_2} &&
 = \int d\mathbf{r}_1^3 \int d\mathbf{r}_2^3
 \Psi^*({\mathbf{p}_1,\mathbf{r}_1;\mathbf{p}_2,\mathbf{r}_2})\\
 && \times 
 V(\mathbf{r}_1,\mathbf{r}_2)
 \Psi({\mathbf{k}_1,\mathbf{r}_1;\mathbf{k}_2,\mathbf{r}_2}),
 \label{Vppkk}
\end{eqnarray}
where $V(\mathbf{r}_1,\mathbf{r}_2)=e^2/(\epsilon|\mathbf{r}_2-\mathbf{r}_1|)$.
The integrand in Eq. (\ref{Vppkk}) contains four terms, but we have $\gamma\sim 0$ for low-energy electrons and holes ($2\hbar vk/\Delta_\infty\ll 1$),
and the terms containing spinor products between 
$\chi^+(x_{1,2},y_{1,2})$ and $\chi^-(x_{1,2},y_{1,2})$ become negligible, see Eqs. (\ref{chi+}) and (\ref{chi-}). The remaining two terms are equal.
Neglecting unimportant phase factors we have
\begin{equation}
 V_{\mathbf{p}_1\mathbf{p}_2\mathbf{k}_1\mathbf{k}_2} \approx
 (2\pi)^2\delta\left(\mathbf{q}-\mathbf{s}\right) V_q,
\end{equation}
where  $\mathbf{q}=\mathbf{p}_1-\mathbf{k}_1$, $\mathbf{s}=\mathbf{k}_2-\mathbf{p}_2$,
and $V_q=2\pi e^2 F_q/(\epsilon q)$ with $F_q$ given by 
\begin{equation}
 F_q= \int\limits_{-\infty}^\infty d z_1\int\limits_{-\infty}^\infty d z_2
 {\mathrm e}^{-q|z_2-z_1|}|\psi_1(z_1)|^2
 |\psi_2(z_2)|^2.
 \label{Fq}
\end{equation}
In the conventional limit of the states localized in the respective layers
we can approximate $|\psi_{1,2}(z_{1,2})|^2=\delta(z_{1,2}\pm d/2)$,
and $F_q={\mathrm e}^{-qd}$. If the wave functions $\psi_{1,2}(z)$ overlap,
then $F_q$ differs strongly from ${\mathrm e}^{-qd}$, as demonstrated in Fig. \ref{fig3}(b).

Introducing $\nu_q=\hbar v q$ we rewrite the gap equation (\ref{MF-4}) as
\begin{equation}
  \int\limits_0^\infty d\nu_q
 \frac{F_{\nu_q/(\hbar v)}}{\sqrt{\Delta_\mathrm{MF}^2 + 
 \left(\sqrt{\nu_q^2+\frac{\Delta_\infty^2}{4}}+\frac{E_1(d)-E_2(d)}{2}\right)^2}}=\frac{2}{r_s},
 \label{MF-5}
\end{equation}
where $r_s=e^2/(\epsilon \hbar v)$.
Figure \ref{fig3}(c) shows solutions of Eq. (\ref{MF-5}) for different semiconductor
bandgaps, with $\Delta_\infty=1.6$ eV being relevant for 2D WSe$_2$ \cite{chaves2020bandgap}.
The velocity $v\sim 10^7$ cm/s is typical for 2D transition metal dichalcogenides \cite{Andor2015},
and the spacer is assumed to be made of $h$-BN with the relative dielectric permittivity of 
$\epsilon=6.9$ \cite{laturia2018dielectric}.
The energy depth is taken to be $E_0= -2$ eV but it could be even deeper
up to $E_0\sim -3$ eV (one-half of the $h$-BN monolayer bandgap) \cite{chaves2020bandgap,zhao2020band}.

Figure \ref{fig3}(c) demonstrates clearly that the wave function overlap 
in $F_q$ is crucial for creating an IEI state at a reasonable $d$.
If the overlap is neglected, then the solutions of Eq. (\ref{MF-5})
shown in Fig. \ref{fig3}(c) by dashed curves exist only at $d<0.5$ nm, i.e. the critical 
$d$ is smaller than the monolayer thickness.
The local maximum of $\Delta_\mathrm{MF}(d)$ occurs at the point 
when $\Delta_\infty \sim E_2(d)-E_1(d)$. It shifts to even 
smaller $d$ when $\Delta_\infty$ increases.  
If the overlap is taken into account, then the critical $d$
shifts towards larger values reaching $2$ nm for $\Delta_\infty=1$ eV, see solid curves in Fig. \ref{fig3}(c).

%Since $F_q$ does not vanish at larger $q$
%The local maximum of $\Delta_\mathrm{MF}(d)$ disappears because so that the integrand contributes in Eq. (\ref{MF-5}) regardless.

Figure \ref{fig3}(d) combines the data shown in Figs. \ref{fig3}(c) and \ref{fig2}(g).
The lower right corner of the phase diagram is the region where IEI state is expected.
In that region, Eq. (\ref{MF-5}) allows for a solution with respect to $\Delta_\mathrm{MF}$,
and, at the same time, the interlayer coherence time, $\tau(d)$, is much longer than
the typical exciton life-time in 2D semiconductors.
In simple terms, the electrons and holes interact strongly but are well separated.
The interlayer EI gradually becomes a single-layer one with decreasing $d$.
There is no sharp border between the two states because there is always a non-zero interlayer hopping probability 
for electrons and holes even though it decays exponentially with increasing $d$.
In contrast, the normal and correlated phases are separated by a sharp border, as shown by solid curve in Fig. \ref{fig3}(d),
because the order parameter equation (\ref{MF-4}) either has a solution or not.

\section{Discussion}
\label{discuss}

\begin{figure}
 \includegraphics[width=\columnwidth]{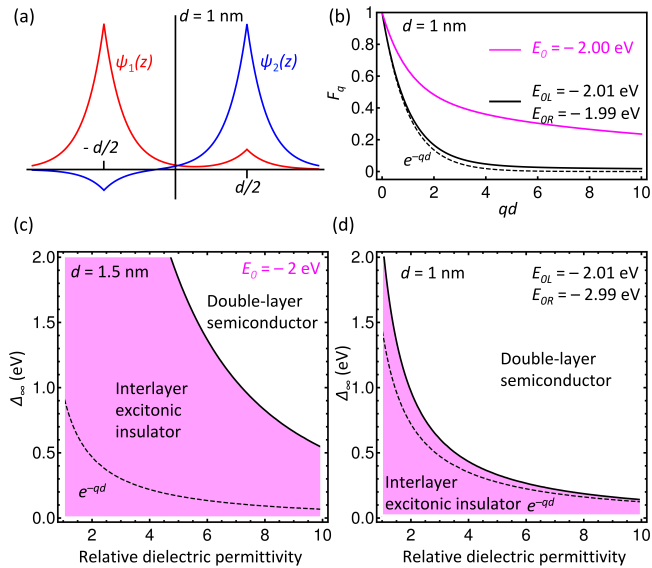}
 \caption{Roles of the double-layer asymmetry, relative dielectric permittivity
 of the interlayer media, and semiconductor bandgap value in formation of the IEI state.
 (a) The wave functions become very asymmetric even though the potential depth 
 difference is small, $(E_{0R}-E_{0L})/(E_{0R}+E_{0L})=0.01$, compare with Fig. \ref{fig1}(c).
 (b) The form-factor approaches $F_q=\mathrm{e}^{-qd}$ when the layer asymmetry increases,
 as if electrons and holes are forced to localize in the respective layers.
 (c) The IEI state is easy to reach in the symmetric double-well configuration 
 with $d\sim 1$ nm separation within any reasonable range of the dielectric permittivity
 and semiconductor bandgap values.
 (d) Reaching the IEI state in an asymmetric double-well is difficult
 but possible for narrow-gap semiconductors and spacers with low dielectric permittivity.
} 
 \label{fig4}
\end{figure}

Having established the crucial role of the interlayer coherence in the IEI state
we now focus on the effects of the double-layer asymmetry, relative dielectric permittivity
of the interlayer media, and the size of a semiconductor bandgap.
The double-layer asymmetry influences the IEI state in two different ways.
On the one hand, the asymmetry reduces the wave function overlap
that results in $F_q\to \mathrm{e}^{-qd}$ making the IEI state harder to reach.
On the other hand, the asymmetry triggers the collapse of electron and hole wave functions 
into their respective layers precluding the interlayer coherence effects \cite{santarsiero2019flea}
and improving electron-hole separation beneficial for the IEI phase.
Note that Eq. (\ref{tau}) estimating $\tau(d)$ makes sense for a symmetric
double-layer only.

To quantify the effect of the double-layer asymmetry we introduce
$E_{0L,R}=-m_0u_{0L,R}^2/(2\hbar^2)$, and instead of Eq. (\ref{U}) we have 
\begin{equation}
 U(z)= - u_{0R} \delta\left(z-\frac{d}{2}\right)- u_{0L} \delta\left(z+\frac{d}{2}\right).
 \label{UA}
\end{equation}
The eigenstates of $H$ can be still written as 
$\Psi_{1,2}^\pm=\psi_{1,2}(z)\chi^\pm(x,y)$ with $\psi_{1,2}(z)$ given by
\begin{equation}
 \psi_{1,2}(z)=\left\{
\begin{array}{ll}
A_{1,2L}\mathrm{e}^{\kappa_{1,2}z}, & z\leq -\frac{d}{2};\\
B_{1,2}\mathrm{e}^{\kappa_{1,2}z} + C_{1,2} \mathrm{e}^{-\kappa_{1,2}z}, & -\frac{d}{2}< z <\frac{d}{2};\\
A_{1,2R}\mathrm{e}^{-\kappa_{1,2}z}, & z\geq \frac{d}{2};
\end{array}
\right.
\end{equation}
where $\kappa_{1,2}$ are the two solutions of the eigenvalue equation given by
\begin{eqnarray}
 \nonumber && \frac{m_0^2 u_{0L} u_{0R}}{\hbar^4} 
 -\left(\kappa - \frac{m_0 u_{0R}}{\hbar^2}\right)\left(\kappa - \frac{m_0 u_{0L}}{\hbar^2}\right){\mathrm e}^{2\kappa d}=0.\\
 \label{kappa-A}
\end{eqnarray}
In contrast to the symmetric case, Eq. (\ref{kappa-A}) does not allow for an explicit solution
but it can be solved numerically \cite{santarsiero2019flea}.
Figure \ref{fig4}(a) shows the states $\psi_{1,2}(z)$ in a slightly asymmetric double-layer.
It is clear that $\psi_1(z)$ tends to collapse into the left layer whereas $\psi_2(z)$ does the same into the right one.
As $\psi_1(z)$ and $\psi_2(z)$ describe, respectively, electron and hole states in our many-body model,
the carriers having opposite charges turn out to be well separated in space.
Hence, the IEI state should exist even at smaller separations.
Figure \ref{fig4}(a) demonstrates, however, that $F_q$ drops significantly even for a slightly asymmetric double-layer and
rapidly approaches the ${\mathrm e}^{-qd}$ form. 
As a consequence, the many-body gap equation has a solution only at a very small $d$, in the region enclosed
by the short-dashed line in Fig. \ref{fig3}(d). Such a small separation does not have physical sense
because it is smaller than the monolayer thickness.
Hence, an asymmetric double-layer structure made of 2D semiconductors with $\Delta_\infty>1$ eV separated by $h$-BN 
is not suitable for creating the IEI state.
A very recent paper \cite{shi2022bilayer}, however, offers a way to suppress interlayer tunneling in semiconducting bilayers without
dielectric spacer.

There are two obvious options to remedy the situation. First, we could reduce $\Delta_\infty$ by using a
narrow gap 2D semiconductor, such as one of X-enes \cite{chaves2020bandgap,PRAoptical,PRBberman} or $1T$-TiS$_2$ \cite{TiSe-1,TiSe-2}.
Second, we could substitute $h$-BN by a dielectric material with a smaller relative dielectric permittivity,
such as silicon dioxide ($\epsilon=3.9$) or polyethylene ($\epsilon=2.25$).
Figures \ref{fig4}(c,d) compare the IEI phase diagrams for the symmetric, Fig. \ref{fig4}(c), and asymmetric, Fig. \ref{fig4}(d), double-layer junctions.
The IEI state is supposed to be stable as long as solution of Eq. (\ref{MF-5}) exists.
The interlayer distance is shown in the respective panels.
The temperature is assumed to be zero.
Note, however, that decreasing semiconductor bandgap might
induce transition to the EI state in each layer separately that would make the desired drag-counterflow setup impossible to implement.
This transition cannot be described by the simplified two-band many-body model employed in
this section.

\section{Outlook}
\label{outro}

The results discussed above suggest that the interlayer overlap between electron and hole wave functions 
potentially facilitates transition into the IEI state.
Figure \ref{fig3}(d) is among the main results of the present paper offering a recipe for the IEI state using known 2D materials.
The ingredients are one insulating and two semiconducting layers.
The semiconducting layers must be identical and possess a direct bandgap of about 1 eV.
A much larger bandgap would require either unrealistically strong interaction or unrealistically small interlayer separation to bring the double-layer
into the IEI state, whereas a much smaller bandgap would convert each layer into the EI state separately.
At the moment, the best choice seems to be 2D TiS$_3$ with the direct bandgap very close to $\Delta_\infty\sim 1$ eV taken in Fig. \ref{fig3}(d).
The bandgap size has been predicted by means of ab-initio calculations \cite{TiS3-abinitio} and confirmed experimentally \cite{TiS3fabrication}.
There is strong anisotropy in electronic structure \cite{TiS3-anisotropic-bands} but it is not able to spoil 
the qualitative applicability of the model proposed.
Moreover, 2D TiS$_3$ can be assembled into heterostructures with other 2D materials, \cite{TiS3heterostructures}
including $h$-BN \cite{TiS3-BN} employed in Fig. \ref{fig3}(d) as a spacer.
As 2D TiS$_3$ and $h$-BN have similar work functions of about 5 eV, and $h$-BN monolayer has a bandgap of about 6 eV,
the resulting band diagram should be similar to that shown in Fig. \ref{fig2}(e) with $E_0\sim -3$ eV.
Having two or three monolayers of $h$-BN as a spacer would bring the electron-hole system into the desired lower-right corner
of the phase diagram in Fig. \ref{fig3}(d).
The black long-dashed curve separating the single-layer EI from the true IEI state also applies to TiS$_3$,
with subpicosecond exciton lifetime \cite{TiS3-exciton-lifetime}.
The main obstacle would be to maintain the layer symmetry in the double-junction.

\acknowledgments

This research is supported by the Ministry of Education, 
Singapore, under its Research Centre of Excellence award to the Institute for Functional Intelligent Materials
(I-FIM, Project No. EDUNC-33-18-279-V12).
I am grateful to the Director's Senior Research Fellowship from the Centre for Advanced 2D Materials at NUS for support
as well as thank Giovanni Vignale, Goki Eda, Alexandra Carvalho, and Aleksandr Rodin for discussions.

\bibliography{XI.bib}

\begin{thebibliography}{46}%
\makeatletter
\providecommand \@ifxundefined [1]{%
 \@ifx{#1\undefined}
}%
\providecommand \@ifnum [1]{%
 \ifnum #1\expandafter \@firstoftwo
 \else \expandafter \@secondoftwo
 \fi
}%
\providecommand \@ifx [1]{%
 \ifx #1\expandafter \@firstoftwo
 \else \expandafter \@secondoftwo
 \fi
}%
\providecommand \natexlab [1]{#1}%
\providecommand \enquote  [1]{``#1''}%
\providecommand \bibnamefont  [1]{#1}%
\providecommand \bibfnamefont [1]{#1}%
\providecommand \citenamefont [1]{#1}%
\providecommand \href@noop [0]{\@secondoftwo}%
\providecommand \href [0]{\begingroup \@sanitize@url \@href}%
\providecommand \@href[1]{\@@startlink{#1}\@@href}%
\providecommand \@@href[1]{\endgroup#1\@@endlink}%
\providecommand \@sanitize@url [0]{\catcode `\\12\catcode `\$12\catcode
  `\&12\catcode `\#12\catcode `\^12\catcode `\_12\catcode `\%12\relax}%
\providecommand \@@startlink[1]{}%
\providecommand \@@endlink[0]{}%
\providecommand \url  [0]{\begingroup\@sanitize@url \@url }%
\providecommand \@url [1]{\endgroup\@href {#1}{\urlprefix }}%
\providecommand \urlprefix  [0]{URL }%
\providecommand \Eprint [0]{\href }%
\providecommand \doibase [0]{http://dx.doi.org/}%
\providecommand \selectlanguage [0]{\@gobble}%
\providecommand \bibinfo  [0]{\@secondoftwo}%
\providecommand \bibfield  [0]{\@secondoftwo}%
\providecommand \translation [1]{[#1]}%
\providecommand \BibitemOpen [0]{}%
\providecommand \bibitemStop [0]{}%
\providecommand \bibitemNoStop [0]{.\EOS\space}%
\providecommand \EOS [0]{\spacefactor3000\relax}%
\providecommand \BibitemShut  [1]{\csname bibitem#1\endcsname}%
\let\auto@bib@innerbib\@empty
%</preamble>
\bibitem [{\citenamefont {Keldysh}\ and\ \citenamefont
  {Kopaev}(1965)}]{keldysh1965possible}%
  \BibitemOpen
  \bibfield  {author} {\bibinfo {author} {\bibfnamefont {L.~V.}\ \bibnamefont
  {Keldysh}}\ and\ \bibinfo {author} {\bibfnamefont {Y.~V.}\ \bibnamefont
  {Kopaev}},\ }\href@noop {} {\bibfield  {journal} {\bibinfo  {journal} {Sov.
  Phys. Solid State, USSR}\ }\textbf {\bibinfo {volume} {6}},\ \bibinfo {pages}
  {2219} (\bibinfo {year} {1965})}\BibitemShut {NoStop}%
\bibitem [{\citenamefont {Cloizeaux}(1965)}]{CLOIZEAUX1965259}%
  \BibitemOpen
  \bibfield  {author} {\bibinfo {author} {\bibfnamefont {J.~D.}\ \bibnamefont
  {Cloizeaux}},\ }\href@noop {} {\bibfield  {journal} {\bibinfo  {journal}
  {Journal of Physics and Chemistry of Solids}\ }\textbf {\bibinfo {volume}
  {26}},\ \bibinfo {pages} {259} (\bibinfo {year} {1965})}\BibitemShut
  {NoStop}%
\bibitem [{\citenamefont {J{\'e}rome}\ \emph {et~al.}(1967)\citenamefont
  {J{\'e}rome}, \citenamefont {Rice},\ and\ \citenamefont
  {Kohn}}]{jerome1967excitonic}%
  \BibitemOpen
  \bibfield  {author} {\bibinfo {author} {\bibfnamefont {D.}~\bibnamefont
  {J{\'e}rome}}, \bibinfo {author} {\bibfnamefont {T.}~\bibnamefont {Rice}}, \
  and\ \bibinfo {author} {\bibfnamefont {W.}~\bibnamefont {Kohn}},\ }\href@noop
  {} {\bibfield  {journal} {\bibinfo  {journal} {Physical Review}\ }\textbf
  {\bibinfo {volume} {158}},\ \bibinfo {pages} {462} (\bibinfo {year}
  {1967})}\BibitemShut {NoStop}%
\bibitem [{\citenamefont {Bucher}\ \emph {et~al.}(1991)\citenamefont {Bucher},
  \citenamefont {Steiner},\ and\ \citenamefont {Wachter}}]{PRL1991bucher}%
  \BibitemOpen
  \bibfield  {author} {\bibinfo {author} {\bibfnamefont {B.}~\bibnamefont
  {Bucher}}, \bibinfo {author} {\bibfnamefont {P.}~\bibnamefont {Steiner}}, \
  and\ \bibinfo {author} {\bibfnamefont {P.}~\bibnamefont {Wachter}},\
  }\href@noop {} {\bibfield  {journal} {\bibinfo  {journal} {Phys. Rev. Lett.}\
  }\textbf {\bibinfo {volume} {67}},\ \bibinfo {pages} {2717} (\bibinfo {year}
  {1991})}\BibitemShut {NoStop}%
\bibitem [{\citenamefont {Cercellier}\ \emph {et~al.}(2007)\citenamefont
  {Cercellier}, \citenamefont {Monney}, \citenamefont {Clerc}, \citenamefont
  {Battaglia}, \citenamefont {Despont}, \citenamefont {Garnier}, \citenamefont
  {Beck}, \citenamefont {Aebi}, \citenamefont {Patthey}, \citenamefont
  {Berger},\ and\ \citenamefont {Forr\'o}}]{PRL2007cercellier}%
  \BibitemOpen
  \bibfield  {author} {\bibinfo {author} {\bibfnamefont {H.}~\bibnamefont
  {Cercellier}}, \bibinfo {author} {\bibfnamefont {C.}~\bibnamefont {Monney}},
  \bibinfo {author} {\bibfnamefont {F.}~\bibnamefont {Clerc}}, \bibinfo
  {author} {\bibfnamefont {C.}~\bibnamefont {Battaglia}}, \bibinfo {author}
  {\bibfnamefont {L.}~\bibnamefont {Despont}}, \bibinfo {author} {\bibfnamefont
  {M.~G.}\ \bibnamefont {Garnier}}, \bibinfo {author} {\bibfnamefont
  {H.}~\bibnamefont {Beck}}, \bibinfo {author} {\bibfnamefont {P.}~\bibnamefont
  {Aebi}}, \bibinfo {author} {\bibfnamefont {L.}~\bibnamefont {Patthey}},
  \bibinfo {author} {\bibfnamefont {H.}~\bibnamefont {Berger}}, \ and\ \bibinfo
  {author} {\bibfnamefont {L.}~\bibnamefont {Forr\'o}},\ }\href@noop {}
  {\bibfield  {journal} {\bibinfo  {journal} {Phys. Rev. Lett.}\ }\textbf
  {\bibinfo {volume} {99}},\ \bibinfo {pages} {146403} (\bibinfo {year}
  {2007})}\BibitemShut {NoStop}%
\bibitem [{\citenamefont {Wakisaka}\ \emph {et~al.}(2009)\citenamefont
  {Wakisaka}, \citenamefont {Sudayama}, \citenamefont {Takubo}, \citenamefont
  {Mizokawa}, \citenamefont {Arita}, \citenamefont {Namatame}, \citenamefont
  {Taniguchi}, \citenamefont {Katayama}, \citenamefont {Nohara},\ and\
  \citenamefont {Takagi}}]{PRL2009wakisaka}%
  \BibitemOpen
  \bibfield  {author} {\bibinfo {author} {\bibfnamefont {Y.}~\bibnamefont
  {Wakisaka}}, \bibinfo {author} {\bibfnamefont {T.}~\bibnamefont {Sudayama}},
  \bibinfo {author} {\bibfnamefont {K.}~\bibnamefont {Takubo}}, \bibinfo
  {author} {\bibfnamefont {T.}~\bibnamefont {Mizokawa}}, \bibinfo {author}
  {\bibfnamefont {M.}~\bibnamefont {Arita}}, \bibinfo {author} {\bibfnamefont
  {H.}~\bibnamefont {Namatame}}, \bibinfo {author} {\bibfnamefont
  {M.}~\bibnamefont {Taniguchi}}, \bibinfo {author} {\bibfnamefont
  {N.}~\bibnamefont {Katayama}}, \bibinfo {author} {\bibfnamefont
  {M.}~\bibnamefont {Nohara}}, \ and\ \bibinfo {author} {\bibfnamefont
  {H.}~\bibnamefont {Takagi}},\ }\href@noop {} {\bibfield  {journal} {\bibinfo
  {journal} {Phys. Rev. Lett.}\ }\textbf {\bibinfo {volume} {103}},\ \bibinfo
  {pages} {026402} (\bibinfo {year} {2009})}\BibitemShut {NoStop}%
\bibitem [{\citenamefont {Lu}\ \emph {et~al.}(2017)\citenamefont {Lu},
  \citenamefont {Kono}, \citenamefont {Larkin}, \citenamefont {Rost},
  \citenamefont {Takayama}, \citenamefont {Boris}, \citenamefont {Keimer},\
  and\ \citenamefont {Takagi}}]{lu2017zero}%
  \BibitemOpen
  \bibfield  {author} {\bibinfo {author} {\bibfnamefont {Y.~F.}\ \bibnamefont
  {Lu}}, \bibinfo {author} {\bibfnamefont {H.}~\bibnamefont {Kono}}, \bibinfo
  {author} {\bibfnamefont {T.~I.}\ \bibnamefont {Larkin}}, \bibinfo {author}
  {\bibfnamefont {A.~W.}\ \bibnamefont {Rost}}, \bibinfo {author}
  {\bibfnamefont {T.}~\bibnamefont {Takayama}}, \bibinfo {author}
  {\bibfnamefont {A.~V.}\ \bibnamefont {Boris}}, \bibinfo {author}
  {\bibfnamefont {B.}~\bibnamefont {Keimer}}, \ and\ \bibinfo {author}
  {\bibfnamefont {H.}~\bibnamefont {Takagi}},\ }\href@noop {} {\bibfield
  {journal} {\bibinfo  {journal} {Nature Commun.}\ }\textbf {\bibinfo {volume}
  {8}},\ \bibinfo {pages} {14408} (\bibinfo {year} {2017})}\BibitemShut
  {NoStop}%
\bibitem [{\citenamefont {Mor}\ \emph {et~al.}(2017)\citenamefont {Mor},
  \citenamefont {Herzog}, \citenamefont {Gole\ifmmode~\check{z}\else
  \v{z}\fi{}}, \citenamefont {Werner}, \citenamefont {Eckstein}, \citenamefont
  {Katayama}, \citenamefont {Nohara}, \citenamefont {Takagi}, \citenamefont
  {Mizokawa}, \citenamefont {Monney},\ and\ \citenamefont
  {St\"ahler}}]{PRL2017gapcontrol}%
  \BibitemOpen
  \bibfield  {author} {\bibinfo {author} {\bibfnamefont {S.}~\bibnamefont
  {Mor}}, \bibinfo {author} {\bibfnamefont {M.}~\bibnamefont {Herzog}},
  \bibinfo {author} {\bibfnamefont {D.}~\bibnamefont
  {Gole\ifmmode~\check{z}\else \v{z}\fi{}}}, \bibinfo {author} {\bibfnamefont
  {P.}~\bibnamefont {Werner}}, \bibinfo {author} {\bibfnamefont
  {M.}~\bibnamefont {Eckstein}}, \bibinfo {author} {\bibfnamefont
  {N.}~\bibnamefont {Katayama}}, \bibinfo {author} {\bibfnamefont
  {M.}~\bibnamefont {Nohara}}, \bibinfo {author} {\bibfnamefont
  {H.}~\bibnamefont {Takagi}}, \bibinfo {author} {\bibfnamefont
  {T.}~\bibnamefont {Mizokawa}}, \bibinfo {author} {\bibfnamefont
  {C.}~\bibnamefont {Monney}}, \ and\ \bibinfo {author} {\bibfnamefont
  {J.}~\bibnamefont {St\"ahler}},\ }\href {\doibase
  10.1103/PhysRevLett.119.086401} {\bibfield  {journal} {\bibinfo  {journal}
  {Phys. Rev. Lett.}\ }\textbf {\bibinfo {volume} {119}},\ \bibinfo {pages}
  {086401} (\bibinfo {year} {2017})}\BibitemShut {NoStop}%
\bibitem [{\citenamefont {Wu}\ \emph {et~al.}(2019)\citenamefont {Wu},
  \citenamefont {Lou}, \citenamefont {Chang}, \citenamefont {Sullivan},
  \citenamefont {Ikhlassi},\ and\ \citenamefont {Du}}]{PRB2019InAs}%
  \BibitemOpen
  \bibfield  {author} {\bibinfo {author} {\bibfnamefont {X.-J.}\ \bibnamefont
  {Wu}}, \bibinfo {author} {\bibfnamefont {W.}~\bibnamefont {Lou}}, \bibinfo
  {author} {\bibfnamefont {K.}~\bibnamefont {Chang}}, \bibinfo {author}
  {\bibfnamefont {G.}~\bibnamefont {Sullivan}}, \bibinfo {author}
  {\bibfnamefont {A.}~\bibnamefont {Ikhlassi}}, \ and\ \bibinfo {author}
  {\bibfnamefont {R.-R.}\ \bibnamefont {Du}},\ }\href {\doibase
  10.1103/PhysRevB.100.165309} {\bibfield  {journal} {\bibinfo  {journal}
  {Phys. Rev. B}\ }\textbf {\bibinfo {volume} {100}},\ \bibinfo {pages}
  {165309} (\bibinfo {year} {2019})}\BibitemShut {NoStop}%
\bibitem [{\citenamefont {Chui}\ \emph {et~al.}(2021)\citenamefont {Chui},
  \citenamefont {Wang},\ and\ \citenamefont {Tanatar}}]{PRB2021QMC}%
  \BibitemOpen
  \bibfield  {author} {\bibinfo {author} {\bibfnamefont {S.~T.}\ \bibnamefont
  {Chui}}, \bibinfo {author} {\bibfnamefont {N.}~\bibnamefont {Wang}}, \ and\
  \bibinfo {author} {\bibfnamefont {B.}~\bibnamefont {Tanatar}},\ }\href
  {\doibase 10.1103/PhysRevB.104.195432} {\bibfield  {journal} {\bibinfo
  {journal} {Phys. Rev. B}\ }\textbf {\bibinfo {volume} {104}},\ \bibinfo
  {pages} {195432} (\bibinfo {year} {2021})}\BibitemShut {NoStop}%
\bibitem [{\citenamefont {Sodemann}\ \emph {et~al.}(2012)\citenamefont
  {Sodemann}, \citenamefont {Pesin},\ and\ \citenamefont
  {MacDonald}}]{PRB2012RPA}%
  \BibitemOpen
  \bibfield  {author} {\bibinfo {author} {\bibfnamefont {I.}~\bibnamefont
  {Sodemann}}, \bibinfo {author} {\bibfnamefont {D.~A.}\ \bibnamefont {Pesin}},
  \ and\ \bibinfo {author} {\bibfnamefont {A.~H.}\ \bibnamefont {MacDonald}},\
  }\href {\doibase 10.1103/PhysRevB.85.195136} {\bibfield  {journal} {\bibinfo
  {journal} {Phys. Rev. B}\ }\textbf {\bibinfo {volume} {85}},\ \bibinfo
  {pages} {195136} (\bibinfo {year} {2012})}\BibitemShut {NoStop}%
\bibitem [{\citenamefont {Brunetti}\ \emph {et~al.}(2018)\citenamefont
  {Brunetti}, \citenamefont {Berman},\ and\ \citenamefont
  {Kezerashvili}}]{Brunetti_2018}%
  \BibitemOpen
  \bibfield  {author} {\bibinfo {author} {\bibfnamefont {M.~N.}\ \bibnamefont
  {Brunetti}}, \bibinfo {author} {\bibfnamefont {O.~L.}\ \bibnamefont
  {Berman}}, \ and\ \bibinfo {author} {\bibfnamefont {R.~Y.}\ \bibnamefont
  {Kezerashvili}},\ }\href {\doibase 10.1088/1361-648x/aabe53} {\bibfield
  {journal} {\bibinfo  {journal} {Journal of Physics: Condensed Matter}\
  }\textbf {\bibinfo {volume} {30}},\ \bibinfo {pages} {225001} (\bibinfo
  {year} {2018})}\BibitemShut {NoStop}%
\bibitem [{\citenamefont {Shi}\ \emph {et~al.}(2022)\citenamefont {Shi},
  \citenamefont {Shih}, \citenamefont {Rhodes}, \citenamefont {Kim},
  \citenamefont {Barmak}, \citenamefont {Watanabe}, \citenamefont {Taniguchi},
  \citenamefont {Papi{\'c}}, \citenamefont {Abanin}, \citenamefont {Hone} \emph
  {et~al.}}]{shi2022bilayer}%
  \BibitemOpen
  \bibfield  {author} {\bibinfo {author} {\bibfnamefont {Q.}~\bibnamefont
  {Shi}}, \bibinfo {author} {\bibfnamefont {E.-M.}\ \bibnamefont {Shih}},
  \bibinfo {author} {\bibfnamefont {D.}~\bibnamefont {Rhodes}}, \bibinfo
  {author} {\bibfnamefont {B.}~\bibnamefont {Kim}}, \bibinfo {author}
  {\bibfnamefont {K.}~\bibnamefont {Barmak}}, \bibinfo {author} {\bibfnamefont
  {K.}~\bibnamefont {Watanabe}}, \bibinfo {author} {\bibfnamefont
  {T.}~\bibnamefont {Taniguchi}}, \bibinfo {author} {\bibfnamefont
  {Z.}~\bibnamefont {Papi{\'c}}}, \bibinfo {author} {\bibfnamefont {D.~A.}\
  \bibnamefont {Abanin}}, \bibinfo {author} {\bibfnamefont {J.}~\bibnamefont
  {Hone}},  \emph {et~al.},\ }\href@noop {} {\bibfield  {journal} {\bibinfo
  {journal} {Nature Nanotechnology}\ }\textbf {\bibinfo {volume} {17}},\
  \bibinfo {pages} {577} (\bibinfo {year} {2022})}\BibitemShut {NoStop}%
\bibitem [{\citenamefont {Du}\ \emph {et~al.}(2017)\citenamefont {Du},
  \citenamefont {Li}, \citenamefont {Lou}, \citenamefont {Sullivan},
  \citenamefont {Chang}, \citenamefont {Kono},\ and\ \citenamefont
  {Du}}]{du2017evidence}%
  \BibitemOpen
  \bibfield  {author} {\bibinfo {author} {\bibfnamefont {L.}~\bibnamefont
  {Du}}, \bibinfo {author} {\bibfnamefont {X.}~\bibnamefont {Li}}, \bibinfo
  {author} {\bibfnamefont {W.}~\bibnamefont {Lou}}, \bibinfo {author}
  {\bibfnamefont {G.}~\bibnamefont {Sullivan}}, \bibinfo {author}
  {\bibfnamefont {K.}~\bibnamefont {Chang}}, \bibinfo {author} {\bibfnamefont
  {J.}~\bibnamefont {Kono}}, \ and\ \bibinfo {author} {\bibfnamefont {R.-R.}\
  \bibnamefont {Du}},\ }\href@noop {} {\bibfield  {journal} {\bibinfo
  {journal} {Nature Commun.}\ }\textbf {\bibinfo {volume} {8}},\ \bibinfo
  {pages} {1971} (\bibinfo {year} {2017})}\BibitemShut {NoStop}%
\bibitem [{\citenamefont {Ma}\ \emph {et~al.}(2021)\citenamefont {Ma},
  \citenamefont {Nguyen}, \citenamefont {Wang}, \citenamefont {Zeng},
  \citenamefont {Watanabe}, \citenamefont {Taniguchi}, \citenamefont
  {MacDonald}, \citenamefont {Mak},\ and\ \citenamefont
  {Shan}}]{ma2021strongly}%
  \BibitemOpen
  \bibfield  {author} {\bibinfo {author} {\bibfnamefont {L.}~\bibnamefont
  {Ma}}, \bibinfo {author} {\bibfnamefont {P.~X.}\ \bibnamefont {Nguyen}},
  \bibinfo {author} {\bibfnamefont {Z.}~\bibnamefont {Wang}}, \bibinfo {author}
  {\bibfnamefont {Y.}~\bibnamefont {Zeng}}, \bibinfo {author} {\bibfnamefont
  {K.}~\bibnamefont {Watanabe}}, \bibinfo {author} {\bibfnamefont
  {T.}~\bibnamefont {Taniguchi}}, \bibinfo {author} {\bibfnamefont {A.~H.}\
  \bibnamefont {MacDonald}}, \bibinfo {author} {\bibfnamefont {K.~F.}\
  \bibnamefont {Mak}}, \ and\ \bibinfo {author} {\bibfnamefont
  {J.}~\bibnamefont {Shan}},\ }\href@noop {} {\bibfield  {journal} {\bibinfo
  {journal} {Nature}\ }\textbf {\bibinfo {volume} {598}},\ \bibinfo {pages}
  {585} (\bibinfo {year} {2021})}\BibitemShut {NoStop}%
\bibitem [{\citenamefont {Zhang}\ \emph {et~al.}(2022)\citenamefont {Zhang},
  \citenamefont {Regan}, \citenamefont {Wang}, \citenamefont {Zhao},
  \citenamefont {Wang}, \citenamefont {Sayyad}, \citenamefont {Yumigeta},
  \citenamefont {Watanabe}, \citenamefont {Taniguchi}, \citenamefont {Tongay}
  \emph {et~al.}}]{zhang2022correlated}%
  \BibitemOpen
  \bibfield  {author} {\bibinfo {author} {\bibfnamefont {Z.}~\bibnamefont
  {Zhang}}, \bibinfo {author} {\bibfnamefont {E.~C.}\ \bibnamefont {Regan}},
  \bibinfo {author} {\bibfnamefont {D.}~\bibnamefont {Wang}}, \bibinfo {author}
  {\bibfnamefont {W.}~\bibnamefont {Zhao}}, \bibinfo {author} {\bibfnamefont
  {S.}~\bibnamefont {Wang}}, \bibinfo {author} {\bibfnamefont {M.}~\bibnamefont
  {Sayyad}}, \bibinfo {author} {\bibfnamefont {K.}~\bibnamefont {Yumigeta}},
  \bibinfo {author} {\bibfnamefont {K.}~\bibnamefont {Watanabe}}, \bibinfo
  {author} {\bibfnamefont {T.}~\bibnamefont {Taniguchi}}, \bibinfo {author}
  {\bibfnamefont {S.}~\bibnamefont {Tongay}},  \emph {et~al.},\ }\href
  {\doibase 10.1038/s41567-022-01702-z} {\bibfield  {journal} {\bibinfo
  {journal} {Nature Physics}\ ,\ \bibinfo {pages} {not available}} (\bibinfo
  {year} {2022})}\BibitemShut {NoStop}%
\bibitem [{\citenamefont {Su}\ and\ \citenamefont
  {MacDonald}(2008)}]{NatPhys2008macdonald}%
  \BibitemOpen
  \bibfield  {author} {\bibinfo {author} {\bibfnamefont {J.-J.}\ \bibnamefont
  {Su}}\ and\ \bibinfo {author} {\bibfnamefont {A.}~\bibnamefont {MacDonald}},\
  }\href@noop {} {\bibfield  {journal} {\bibinfo  {journal} {Nature Physics}\
  }\textbf {\bibinfo {volume} {4}},\ \bibinfo {pages} {799} (\bibinfo {year}
  {2008})}\BibitemShut {NoStop}%
\bibitem [{\citenamefont {Xie}\ and\ \citenamefont
  {MacDonald}(2018)}]{macdonald-interlayer-tunnel}%
  \BibitemOpen
  \bibfield  {author} {\bibinfo {author} {\bibfnamefont {M.}~\bibnamefont
  {Xie}}\ and\ \bibinfo {author} {\bibfnamefont {A.~H.}\ \bibnamefont
  {MacDonald}},\ }\href {\doibase 10.1103/PhysRevLett.121.067702} {\bibfield
  {journal} {\bibinfo  {journal} {Phys. Rev. Lett.}\ }\textbf {\bibinfo
  {volume} {121}},\ \bibinfo {pages} {067702} (\bibinfo {year}
  {2018})}\BibitemShut {NoStop}%
\bibitem [{\citenamefont {Eisenstein}\ and\ \citenamefont
  {MacDonald}(2004)}]{eisenstein2004bose}%
  \BibitemOpen
  \bibfield  {author} {\bibinfo {author} {\bibfnamefont {J.}~\bibnamefont
  {Eisenstein}}\ and\ \bibinfo {author} {\bibfnamefont {A.}~\bibnamefont
  {MacDonald}},\ }\href@noop {} {\bibfield  {journal} {\bibinfo  {journal}
  {Nature}\ }\textbf {\bibinfo {volume} {432}},\ \bibinfo {pages} {691}
  (\bibinfo {year} {2004})}\BibitemShut {NoStop}%
\bibitem [{\citenamefont {Vignale}\ and\ \citenamefont
  {MacDonald}(1996)}]{vignale-macdonald-1996}%
  \BibitemOpen
  \bibfield  {author} {\bibinfo {author} {\bibfnamefont {G.}~\bibnamefont
  {Vignale}}\ and\ \bibinfo {author} {\bibfnamefont {A.~H.}\ \bibnamefont
  {MacDonald}},\ }\href {\doibase 10.1103/PhysRevLett.76.2786} {\bibfield
  {journal} {\bibinfo  {journal} {Phys. Rev. Lett.}\ }\textbf {\bibinfo
  {volume} {76}},\ \bibinfo {pages} {2786} (\bibinfo {year}
  {1996})}\BibitemShut {NoStop}%
\bibitem [{\citenamefont {Trushin}(2019)}]{TightlyEx2019}%
  \BibitemOpen
  \bibfield  {author} {\bibinfo {author} {\bibfnamefont {M.}~\bibnamefont
  {Trushin}},\ }\href {\doibase 10.1103/PhysRevB.99.205307} {\bibfield
  {journal} {\bibinfo  {journal} {Phys. Rev. B}\ }\textbf {\bibinfo {volume}
  {99}},\ \bibinfo {pages} {205307} (\bibinfo {year} {2019})}\BibitemShut
  {NoStop}%
\bibitem [{\citenamefont {Jia}\ \emph {et~al.}(2022)\citenamefont {Jia},
  \citenamefont {Wang}, \citenamefont {Chiu}, \citenamefont {Song},
  \citenamefont {Yu}, \citenamefont {J{\"a}ck}, \citenamefont {Lei},
  \citenamefont {Klemenz}, \citenamefont {Cevallos}, \citenamefont {Onyszczak}
  \emph {et~al.}}]{jia2022evidence}%
  \BibitemOpen
  \bibfield  {author} {\bibinfo {author} {\bibfnamefont {Y.}~\bibnamefont
  {Jia}}, \bibinfo {author} {\bibfnamefont {P.}~\bibnamefont {Wang}}, \bibinfo
  {author} {\bibfnamefont {C.-L.}\ \bibnamefont {Chiu}}, \bibinfo {author}
  {\bibfnamefont {Z.}~\bibnamefont {Song}}, \bibinfo {author} {\bibfnamefont
  {G.}~\bibnamefont {Yu}}, \bibinfo {author} {\bibfnamefont {B.}~\bibnamefont
  {J{\"a}ck}}, \bibinfo {author} {\bibfnamefont {S.}~\bibnamefont {Lei}},
  \bibinfo {author} {\bibfnamefont {S.}~\bibnamefont {Klemenz}}, \bibinfo
  {author} {\bibfnamefont {F.~A.}\ \bibnamefont {Cevallos}}, \bibinfo {author}
  {\bibfnamefont {M.}~\bibnamefont {Onyszczak}},  \emph {et~al.},\ }\href@noop
  {} {\bibfield  {journal} {\bibinfo  {journal} {Nature Physics}\ }\textbf
  {\bibinfo {volume} {18}},\ \bibinfo {pages} {87} (\bibinfo {year}
  {2022})}\BibitemShut {NoStop}%
\bibitem [{\citenamefont {Kohn}\ and\ \citenamefont
  {Sherrington}(1970)}]{Kohn1970}%
  \BibitemOpen
  \bibfield  {author} {\bibinfo {author} {\bibfnamefont {W.}~\bibnamefont
  {Kohn}}\ and\ \bibinfo {author} {\bibfnamefont {D.}~\bibnamefont
  {Sherrington}},\ }\href {\doibase 10.1103/RevModPhys.42.1} {\bibfield
  {journal} {\bibinfo  {journal} {Rev. Mod. Phys.}\ }\textbf {\bibinfo {volume}
  {42}},\ \bibinfo {pages} {1} (\bibinfo {year} {1970})}\BibitemShut {NoStop}%
\bibitem [{\citenamefont {Zhu}\ \emph {et~al.}(1995)\citenamefont {Zhu},
  \citenamefont {Littlewood}, \citenamefont {Hybertsen},\ and\ \citenamefont
  {Rice}}]{Littlewood1995}%
  \BibitemOpen
  \bibfield  {author} {\bibinfo {author} {\bibfnamefont {X.}~\bibnamefont
  {Zhu}}, \bibinfo {author} {\bibfnamefont {P.~B.}\ \bibnamefont {Littlewood}},
  \bibinfo {author} {\bibfnamefont {M.~S.}\ \bibnamefont {Hybertsen}}, \ and\
  \bibinfo {author} {\bibfnamefont {T.~M.}\ \bibnamefont {Rice}},\ }\href
  {\doibase 10.1103/PhysRevLett.74.1633} {\bibfield  {journal} {\bibinfo
  {journal} {Phys. Rev. Lett.}\ }\textbf {\bibinfo {volume} {74}},\ \bibinfo
  {pages} {1633} (\bibinfo {year} {1995})}\BibitemShut {NoStop}%
\bibitem [{\citenamefont {Conti}\ \emph {et~al.}(1998)\citenamefont {Conti},
  \citenamefont {Vignale},\ and\ \citenamefont
  {MacDonald}}]{vignale-macdonald-1998}%
  \BibitemOpen
  \bibfield  {author} {\bibinfo {author} {\bibfnamefont {S.}~\bibnamefont
  {Conti}}, \bibinfo {author} {\bibfnamefont {G.}~\bibnamefont {Vignale}}, \
  and\ \bibinfo {author} {\bibfnamefont {A.~H.}\ \bibnamefont {MacDonald}},\
  }\href {\doibase 10.1103/PhysRevB.57.R6846} {\bibfield  {journal} {\bibinfo
  {journal} {Phys. Rev. B}\ }\textbf {\bibinfo {volume} {57}},\ \bibinfo
  {pages} {R6846} (\bibinfo {year} {1998})}\BibitemShut {NoStop}%
\bibitem [{\citenamefont {Korm{\'{a}}nyos}\ \emph {et~al.}(2015)\citenamefont
  {Korm{\'{a}}nyos}, \citenamefont {Burkard}, \citenamefont {Gmitra},
  \citenamefont {Fabian}, \citenamefont {Z{\'{o}}lyomi}, \citenamefont
  {Drummond},\ and\ \citenamefont {Fal'ko}}]{Andor2015}%
  \BibitemOpen
  \bibfield  {author} {\bibinfo {author} {\bibfnamefont {A.}~\bibnamefont
  {Korm{\'{a}}nyos}}, \bibinfo {author} {\bibfnamefont {G.}~\bibnamefont
  {Burkard}}, \bibinfo {author} {\bibfnamefont {M.}~\bibnamefont {Gmitra}},
  \bibinfo {author} {\bibfnamefont {J.}~\bibnamefont {Fabian}}, \bibinfo
  {author} {\bibfnamefont {V.}~\bibnamefont {Z{\'{o}}lyomi}}, \bibinfo {author}
  {\bibfnamefont {N.~D.}\ \bibnamefont {Drummond}}, \ and\ \bibinfo {author}
  {\bibfnamefont {V.}~\bibnamefont {Fal'ko}},\ }\href {\doibase
  10.1088/2053-1583/2/2/022001} {\bibfield  {journal} {\bibinfo  {journal} {2D
  Materials}\ }\textbf {\bibinfo {volume} {2}},\ \bibinfo {pages} {022001}
  (\bibinfo {year} {2015})}\BibitemShut {NoStop}%
\bibitem [{\citenamefont {Xiao}\ \emph {et~al.}(2012)\citenamefont {Xiao},
  \citenamefont {Liu}, \citenamefont {Feng}, \citenamefont {Xu},\ and\
  \citenamefont {Yao}}]{PRL2012xiao}%
  \BibitemOpen
  \bibfield  {author} {\bibinfo {author} {\bibfnamefont {D.}~\bibnamefont
  {Xiao}}, \bibinfo {author} {\bibfnamefont {G.-B.}\ \bibnamefont {Liu}},
  \bibinfo {author} {\bibfnamefont {W.}~\bibnamefont {Feng}}, \bibinfo {author}
  {\bibfnamefont {X.}~\bibnamefont {Xu}}, \ and\ \bibinfo {author}
  {\bibfnamefont {W.}~\bibnamefont {Yao}},\ }\href {\doibase
  10.1103/PhysRevLett.108.196802} {\bibfield  {journal} {\bibinfo  {journal}
  {Phys. Rev. Lett.}\ }\textbf {\bibinfo {volume} {108}},\ \bibinfo {pages}
  {196802} (\bibinfo {year} {2012})}\BibitemShut {NoStop}%
\bibitem [{\citenamefont {Santarsiero}\ and\ \citenamefont
  {Gori}(2019)}]{santarsiero2019flea}%
  \BibitemOpen
  \bibfield  {author} {\bibinfo {author} {\bibfnamefont {M.}~\bibnamefont
  {Santarsiero}}\ and\ \bibinfo {author} {\bibfnamefont {F.}~\bibnamefont
  {Gori}},\ }\href@noop {} {\bibfield  {journal} {\bibinfo  {journal} {European
  Journal of Physics}\ }\textbf {\bibinfo {volume} {40}},\ \bibinfo {pages}
  {055402} (\bibinfo {year} {2019})}\BibitemShut {NoStop}%
\bibitem [{\citenamefont {Ahmed}\ \emph {et~al.}(2016)\citenamefont {Ahmed},
  \citenamefont {Kumar}, \citenamefont {Sharma},\ and\ \citenamefont
  {Sharma}}]{ahmed2016revisiting}%
  \BibitemOpen
  \bibfield  {author} {\bibinfo {author} {\bibfnamefont {Z.}~\bibnamefont
  {Ahmed}}, \bibinfo {author} {\bibfnamefont {S.}~\bibnamefont {Kumar}},
  \bibinfo {author} {\bibfnamefont {M.}~\bibnamefont {Sharma}}, \ and\ \bibinfo
  {author} {\bibfnamefont {V.}~\bibnamefont {Sharma}},\ }\href@noop {}
  {\bibfield  {journal} {\bibinfo  {journal} {European Journal of Physics}\
  }\textbf {\bibinfo {volume} {37}},\ \bibinfo {pages} {045406} (\bibinfo
  {year} {2016})}\BibitemShut {NoStop}%
\bibitem [{\citenamefont {{\"O}zcelik}\ \emph {et~al.}(2016)\citenamefont
  {{\"O}zcelik}, \citenamefont {Azadani}, \citenamefont {Yang}, \citenamefont
  {Koester},\ and\ \citenamefont {Low}}]{ozcelik2016band}%
  \BibitemOpen
  \bibfield  {author} {\bibinfo {author} {\bibfnamefont {V.~O.}\ \bibnamefont
  {{\"O}zcelik}}, \bibinfo {author} {\bibfnamefont {J.~G.}\ \bibnamefont
  {Azadani}}, \bibinfo {author} {\bibfnamefont {C.}~\bibnamefont {Yang}},
  \bibinfo {author} {\bibfnamefont {S.~J.}\ \bibnamefont {Koester}}, \ and\
  \bibinfo {author} {\bibfnamefont {T.}~\bibnamefont {Low}},\ }\href@noop {}
  {\bibfield  {journal} {\bibinfo  {journal} {Physical Review B}\ }\textbf
  {\bibinfo {volume} {94}},\ \bibinfo {pages} {035125} (\bibinfo {year}
  {2016})}\BibitemShut {NoStop}%
\bibitem [{\citenamefont {Wang}\ \emph {et~al.}(2015)\citenamefont {Wang},
  \citenamefont {Zhang},\ and\ \citenamefont {Rana}}]{MoS2-2015}%
  \BibitemOpen
  \bibfield  {author} {\bibinfo {author} {\bibfnamefont {H.}~\bibnamefont
  {Wang}}, \bibinfo {author} {\bibfnamefont {C.}~\bibnamefont {Zhang}}, \ and\
  \bibinfo {author} {\bibfnamefont {F.}~\bibnamefont {Rana}},\ }\href {\doibase
  10.1021/nl503636c} {\bibfield  {journal} {\bibinfo  {journal} {Nano Letters}\
  }\textbf {\bibinfo {volume} {15}},\ \bibinfo {pages} {339} (\bibinfo {year}
  {2015})}\BibitemShut {NoStop}%
\bibitem [{\citenamefont {Yuan}\ and\ \citenamefont {Huang}(2015)}]{WS2-2015}%
  \BibitemOpen
  \bibfield  {author} {\bibinfo {author} {\bibfnamefont {L.}~\bibnamefont
  {Yuan}}\ and\ \bibinfo {author} {\bibfnamefont {L.}~\bibnamefont {Huang}},\
  }\href {\doibase 10.1039/C5NR00383K} {\bibfield  {journal} {\bibinfo
  {journal} {Nanoscale}\ }\textbf {\bibinfo {volume} {7}},\ \bibinfo {pages}
  {7402} (\bibinfo {year} {2015})}\BibitemShut {NoStop}%
\bibitem [{\citenamefont {Littlewood}\ \emph {et~al.}(2004)\citenamefont
  {Littlewood}, \citenamefont {Eastham}, \citenamefont {Keeling}, \citenamefont
  {Marchetti}, \citenamefont {Simons},\ and\ \citenamefont
  {Szymanska}}]{Littlewood2004}%
  \BibitemOpen
  \bibfield  {author} {\bibinfo {author} {\bibfnamefont {P.~B.}\ \bibnamefont
  {Littlewood}}, \bibinfo {author} {\bibfnamefont {P.~R.}\ \bibnamefont
  {Eastham}}, \bibinfo {author} {\bibfnamefont {J.~M.~J.}\ \bibnamefont
  {Keeling}}, \bibinfo {author} {\bibfnamefont {F.~M.}\ \bibnamefont
  {Marchetti}}, \bibinfo {author} {\bibfnamefont {B.~D.}\ \bibnamefont
  {Simons}}, \ and\ \bibinfo {author} {\bibfnamefont {M.~H.}\ \bibnamefont
  {Szymanska}},\ }\href {\doibase 10.1088/0953-8984/16/35/003} {\bibfield
  {journal} {\bibinfo  {journal} {Journal of Physics: Condensed Matter}\
  }\textbf {\bibinfo {volume} {16}},\ \bibinfo {pages} {S3597} (\bibinfo {year}
  {2004})}\BibitemShut {NoStop}%
\bibitem [{\citenamefont {Chaves}\ \emph {et~al.}(2020)\citenamefont {Chaves},
  \citenamefont {Azadani}, \citenamefont {Alsalman}, \citenamefont {Da~Costa},
  \citenamefont {Frisenda}, \citenamefont {Chaves}, \citenamefont {Song},
  \citenamefont {Kim}, \citenamefont {He}, \citenamefont {Zhou} \emph
  {et~al.}}]{chaves2020bandgap}%
  \BibitemOpen
  \bibfield  {author} {\bibinfo {author} {\bibfnamefont {A.}~\bibnamefont
  {Chaves}}, \bibinfo {author} {\bibfnamefont {J.}~\bibnamefont {Azadani}},
  \bibinfo {author} {\bibfnamefont {H.}~\bibnamefont {Alsalman}}, \bibinfo
  {author} {\bibfnamefont {D.}~\bibnamefont {Da~Costa}}, \bibinfo {author}
  {\bibfnamefont {R.}~\bibnamefont {Frisenda}}, \bibinfo {author}
  {\bibfnamefont {A.}~\bibnamefont {Chaves}}, \bibinfo {author} {\bibfnamefont
  {S.~H.}\ \bibnamefont {Song}}, \bibinfo {author} {\bibfnamefont
  {Y.}~\bibnamefont {Kim}}, \bibinfo {author} {\bibfnamefont {D.}~\bibnamefont
  {He}}, \bibinfo {author} {\bibfnamefont {J.}~\bibnamefont {Zhou}},  \emph
  {et~al.},\ }\href@noop {} {\bibfield  {journal} {\bibinfo  {journal} {npj 2D
  Materials and Applications}\ }\textbf {\bibinfo {volume} {4}},\ \bibinfo
  {pages} {1} (\bibinfo {year} {2020})}\BibitemShut {NoStop}%
\bibitem [{\citenamefont {Laturia}\ \emph {et~al.}(2018)\citenamefont
  {Laturia}, \citenamefont {Van~de Put},\ and\ \citenamefont
  {Vandenberghe}}]{laturia2018dielectric}%
  \BibitemOpen
  \bibfield  {author} {\bibinfo {author} {\bibfnamefont {A.}~\bibnamefont
  {Laturia}}, \bibinfo {author} {\bibfnamefont {M.~L.}\ \bibnamefont {Van~de
  Put}}, \ and\ \bibinfo {author} {\bibfnamefont {W.~G.}\ \bibnamefont
  {Vandenberghe}},\ }\href@noop {} {\bibfield  {journal} {\bibinfo  {journal}
  {npj 2D Materials and Applications}\ }\textbf {\bibinfo {volume} {2}},\
  \bibinfo {pages} {1} (\bibinfo {year} {2018})}\BibitemShut {NoStop}%
\bibitem [{\citenamefont {Zhao}\ \emph {et~al.}(2020)\citenamefont {Zhao},
  \citenamefont {Wang}, \citenamefont {Xu}, \citenamefont {Ma}, \citenamefont
  {Li}, \citenamefont {Yang}, \citenamefont {Liu}, \citenamefont {Yang} \emph
  {et~al.}}]{zhao2020band}%
  \BibitemOpen
  \bibfield  {author} {\bibinfo {author} {\bibfnamefont {G.}~\bibnamefont
  {Zhao}}, \bibinfo {author} {\bibfnamefont {J.}~\bibnamefont {Wang}}, \bibinfo
  {author} {\bibfnamefont {Y.}~\bibnamefont {Xu}}, \bibinfo {author}
  {\bibfnamefont {Z.}~\bibnamefont {Ma}}, \bibinfo {author} {\bibfnamefont
  {X.}~\bibnamefont {Li}}, \bibinfo {author} {\bibfnamefont {W.}~\bibnamefont
  {Yang}}, \bibinfo {author} {\bibfnamefont {G.}~\bibnamefont {Liu}}, \bibinfo
  {author} {\bibfnamefont {J.}~\bibnamefont {Yang}},  \emph {et~al.},\
  }\href@noop {} {\bibfield  {journal} {\bibinfo  {journal} {Journal of Alloys
  and Compounds}\ }\textbf {\bibinfo {volume} {834}},\ \bibinfo {pages}
  {155108} (\bibinfo {year} {2020})}\BibitemShut {NoStop}%
\bibitem [{\citenamefont {Zheng}\ \emph {et~al.}(2020)\citenamefont {Zheng},
  \citenamefont {Xiang}, \citenamefont {Li}, \citenamefont {Yuan},
  \citenamefont {Chi},\ and\ \citenamefont {Guo}}]{PRAoptical}%
  \BibitemOpen
  \bibfield  {author} {\bibinfo {author} {\bibfnamefont {J.}~\bibnamefont
  {Zheng}}, \bibinfo {author} {\bibfnamefont {Y.}~\bibnamefont {Xiang}},
  \bibinfo {author} {\bibfnamefont {C.}~\bibnamefont {Li}}, \bibinfo {author}
  {\bibfnamefont {R.}~\bibnamefont {Yuan}}, \bibinfo {author} {\bibfnamefont
  {F.}~\bibnamefont {Chi}}, \ and\ \bibinfo {author} {\bibfnamefont
  {Y.}~\bibnamefont {Guo}},\ }\href {\doibase 10.1103/PhysRevApplied.14.034027}
  {\bibfield  {journal} {\bibinfo  {journal} {Phys. Rev. Applied}\ }\textbf
  {\bibinfo {volume} {14}},\ \bibinfo {pages} {034027} (\bibinfo {year}
  {2020})}\BibitemShut {NoStop}%
\bibitem [{\citenamefont {Brunetti}\ \emph {et~al.}(2019)\citenamefont
  {Brunetti}, \citenamefont {Berman},\ and\ \citenamefont
  {Kezerashvili}}]{PRBberman}%
  \BibitemOpen
  \bibfield  {author} {\bibinfo {author} {\bibfnamefont {M.~N.}\ \bibnamefont
  {Brunetti}}, \bibinfo {author} {\bibfnamefont {O.~L.}\ \bibnamefont
  {Berman}}, \ and\ \bibinfo {author} {\bibfnamefont {R.~Y.}\ \bibnamefont
  {Kezerashvili}},\ }\href {\doibase 10.1103/PhysRevB.99.195417} {\bibfield
  {journal} {\bibinfo  {journal} {Phys. Rev. B}\ }\textbf {\bibinfo {volume}
  {99}},\ \bibinfo {pages} {195417} (\bibinfo {year} {2019})}\BibitemShut
  {NoStop}%
\bibitem [{\citenamefont {Wei}\ \emph {et~al.}(2017)\citenamefont {Wei},
  \citenamefont {Lu}, \citenamefont {Xiao}, \citenamefont {Lv}, \citenamefont
  {Tong}, \citenamefont {Song},\ and\ \citenamefont {Sun}}]{TiSe-1}%
  \BibitemOpen
  \bibfield  {author} {\bibinfo {author} {\bibfnamefont {M.~J.}\ \bibnamefont
  {Wei}}, \bibinfo {author} {\bibfnamefont {W.~J.}\ \bibnamefont {Lu}},
  \bibinfo {author} {\bibfnamefont {R.~C.}\ \bibnamefont {Xiao}}, \bibinfo
  {author} {\bibfnamefont {H.~Y.}\ \bibnamefont {Lv}}, \bibinfo {author}
  {\bibfnamefont {P.}~\bibnamefont {Tong}}, \bibinfo {author} {\bibfnamefont
  {W.~H.}\ \bibnamefont {Song}}, \ and\ \bibinfo {author} {\bibfnamefont
  {Y.~P.}\ \bibnamefont {Sun}},\ }\href {\doibase 10.1103/PhysRevB.96.165404}
  {\bibfield  {journal} {\bibinfo  {journal} {Phys. Rev. B}\ }\textbf {\bibinfo
  {volume} {96}},\ \bibinfo {pages} {165404} (\bibinfo {year}
  {2017})}\BibitemShut {NoStop}%
\bibitem [{\citenamefont {Singh}\ \emph {et~al.}(2017)\citenamefont {Singh},
  \citenamefont {Hsu}, \citenamefont {Tsai}, \citenamefont {Pereira},\ and\
  \citenamefont {Lin}}]{TiSe-2}%
  \BibitemOpen
  \bibfield  {author} {\bibinfo {author} {\bibfnamefont {B.}~\bibnamefont
  {Singh}}, \bibinfo {author} {\bibfnamefont {C.-H.}\ \bibnamefont {Hsu}},
  \bibinfo {author} {\bibfnamefont {W.-F.}\ \bibnamefont {Tsai}}, \bibinfo
  {author} {\bibfnamefont {V.~M.}\ \bibnamefont {Pereira}}, \ and\ \bibinfo
  {author} {\bibfnamefont {H.}~\bibnamefont {Lin}},\ }\href {\doibase
  10.1103/PhysRevB.95.245136} {\bibfield  {journal} {\bibinfo  {journal} {Phys.
  Rev. B}\ }\textbf {\bibinfo {volume} {95}},\ \bibinfo {pages} {245136}
  (\bibinfo {year} {2017})}\BibitemShut {NoStop}%
\bibitem [{\citenamefont {Dai}\ and\ \citenamefont
  {Zeng}(2015)}]{TiS3-abinitio}%
  \BibitemOpen
  \bibfield  {author} {\bibinfo {author} {\bibfnamefont {J.}~\bibnamefont
  {Dai}}\ and\ \bibinfo {author} {\bibfnamefont {X.~C.}\ \bibnamefont {Zeng}},\
  }\href@noop {} {\bibfield  {journal} {\bibinfo  {journal} {Angewandte
  Chemie}\ }\textbf {\bibinfo {volume} {127}},\ \bibinfo {pages} {7682}
  (\bibinfo {year} {2015})}\BibitemShut {NoStop}%
\bibitem [{\citenamefont {Island}\ \emph {et~al.}(2015)\citenamefont {Island},
  \citenamefont {Barawi}, \citenamefont {Biele}, \citenamefont {Almaz{\'a}n},
  \citenamefont {Clamagirand}, \citenamefont {Ares}, \citenamefont
  {S{\'a}nchez}, \citenamefont {Van Der~Zant}, \citenamefont {{\'A}lvarez},
  \citenamefont {D'Agosta} \emph {et~al.}}]{TiS3fabrication}%
  \BibitemOpen
  \bibfield  {author} {\bibinfo {author} {\bibfnamefont {J.~O.}\ \bibnamefont
  {Island}}, \bibinfo {author} {\bibfnamefont {M.}~\bibnamefont {Barawi}},
  \bibinfo {author} {\bibfnamefont {R.}~\bibnamefont {Biele}}, \bibinfo
  {author} {\bibfnamefont {A.}~\bibnamefont {Almaz{\'a}n}}, \bibinfo {author}
  {\bibfnamefont {J.~M.}\ \bibnamefont {Clamagirand}}, \bibinfo {author}
  {\bibfnamefont {J.~R.}\ \bibnamefont {Ares}}, \bibinfo {author}
  {\bibfnamefont {C.}~\bibnamefont {S{\'a}nchez}}, \bibinfo {author}
  {\bibfnamefont {H.~S.}\ \bibnamefont {Van Der~Zant}}, \bibinfo {author}
  {\bibfnamefont {J.~V.}\ \bibnamefont {{\'A}lvarez}}, \bibinfo {author}
  {\bibfnamefont {R.}~\bibnamefont {D'Agosta}},  \emph {et~al.},\ }\href@noop
  {} {\bibfield  {journal} {\bibinfo  {journal} {Advanced Materials}\ }\textbf
  {\bibinfo {volume} {27}},\ \bibinfo {pages} {2595} (\bibinfo {year}
  {2015})}\BibitemShut {NoStop}%
\bibitem [{\citenamefont {Silva-Guill{\'e}n}\ \emph {et~al.}(2017)\citenamefont
  {Silva-Guill{\'e}n}, \citenamefont {Canadell}, \citenamefont {Ordej{\'o}n},
  \citenamefont {Guinea},\ and\ \citenamefont
  {Roldan}}]{TiS3-anisotropic-bands}%
  \BibitemOpen
  \bibfield  {author} {\bibinfo {author} {\bibfnamefont {J.~A.}\ \bibnamefont
  {Silva-Guill{\'e}n}}, \bibinfo {author} {\bibfnamefont {E.}~\bibnamefont
  {Canadell}}, \bibinfo {author} {\bibfnamefont {P.}~\bibnamefont
  {Ordej{\'o}n}}, \bibinfo {author} {\bibfnamefont {F.}~\bibnamefont {Guinea}},
  \ and\ \bibinfo {author} {\bibfnamefont {R.}~\bibnamefont {Roldan}},\
  }\href@noop {} {\bibfield  {journal} {\bibinfo  {journal} {2D Materials}\
  }\textbf {\bibinfo {volume} {4}},\ \bibinfo {pages} {025085} (\bibinfo {year}
  {2017})}\BibitemShut {NoStop}%
\bibitem [{\citenamefont {Liu}\ \emph {et~al.}(2018)\citenamefont {Liu},
  \citenamefont {Guo}, \citenamefont {Wang},\ and\ \citenamefont
  {Wang}}]{TiS3heterostructures}%
  \BibitemOpen
  \bibfield  {author} {\bibinfo {author} {\bibfnamefont {J.}~\bibnamefont
  {Liu}}, \bibinfo {author} {\bibfnamefont {Y.}~\bibnamefont {Guo}}, \bibinfo
  {author} {\bibfnamefont {F.~Q.}\ \bibnamefont {Wang}}, \ and\ \bibinfo
  {author} {\bibfnamefont {Q.}~\bibnamefont {Wang}},\ }\href@noop {} {\bibfield
   {journal} {\bibinfo  {journal} {Nanoscale}\ }\textbf {\bibinfo {volume}
  {10}},\ \bibinfo {pages} {807} (\bibinfo {year} {2018})}\BibitemShut
  {NoStop}%
\bibitem [{\citenamefont {Papadopoulos}\ \emph {et~al.}(2019)\citenamefont
  {Papadopoulos}, \citenamefont {Flores}, \citenamefont {Watanabe},
  \citenamefont {Taniguchi}, \citenamefont {Ares}, \citenamefont {Sanchez},
  \citenamefont {Ferrer}, \citenamefont {Castellanos-Gomez}, \citenamefont
  {Steele},\ and\ \citenamefont {Van Der~Zant}}]{TiS3-BN}%
  \BibitemOpen
  \bibfield  {author} {\bibinfo {author} {\bibfnamefont {N.}~\bibnamefont
  {Papadopoulos}}, \bibinfo {author} {\bibfnamefont {E.}~\bibnamefont
  {Flores}}, \bibinfo {author} {\bibfnamefont {K.}~\bibnamefont {Watanabe}},
  \bibinfo {author} {\bibfnamefont {T.}~\bibnamefont {Taniguchi}}, \bibinfo
  {author} {\bibfnamefont {J.~R.}\ \bibnamefont {Ares}}, \bibinfo {author}
  {\bibfnamefont {C.}~\bibnamefont {Sanchez}}, \bibinfo {author} {\bibfnamefont
  {I.~J.}\ \bibnamefont {Ferrer}}, \bibinfo {author} {\bibfnamefont
  {A.}~\bibnamefont {Castellanos-Gomez}}, \bibinfo {author} {\bibfnamefont
  {G.~A.}\ \bibnamefont {Steele}}, \ and\ \bibinfo {author} {\bibfnamefont
  {H.~S.}\ \bibnamefont {Van Der~Zant}},\ }\href@noop {} {\bibfield  {journal}
  {\bibinfo  {journal} {2D Materials}\ }\textbf {\bibinfo {volume} {7}},\
  \bibinfo {pages} {015009} (\bibinfo {year} {2019})}\BibitemShut {NoStop}%
\bibitem [{\citenamefont {Cui}\ \emph {et~al.}(2016)\citenamefont {Cui},
  \citenamefont {Lipatov}, \citenamefont {Wilt}, \citenamefont {Bellus},
  \citenamefont {Zeng}, \citenamefont {Wu}, \citenamefont {Sinitskii},\ and\
  \citenamefont {Zhao}}]{TiS3-exciton-lifetime}%
  \BibitemOpen
  \bibfield  {author} {\bibinfo {author} {\bibfnamefont {Q.}~\bibnamefont
  {Cui}}, \bibinfo {author} {\bibfnamefont {A.}~\bibnamefont {Lipatov}},
  \bibinfo {author} {\bibfnamefont {J.~S.}\ \bibnamefont {Wilt}}, \bibinfo
  {author} {\bibfnamefont {M.~Z.}\ \bibnamefont {Bellus}}, \bibinfo {author}
  {\bibfnamefont {X.~C.}\ \bibnamefont {Zeng}}, \bibinfo {author}
  {\bibfnamefont {J.}~\bibnamefont {Wu}}, \bibinfo {author} {\bibfnamefont
  {A.}~\bibnamefont {Sinitskii}}, \ and\ \bibinfo {author} {\bibfnamefont
  {H.}~\bibnamefont {Zhao}},\ }\href@noop {} {\bibfield  {journal} {\bibinfo
  {journal} {ACS Applied Materials \& Interfaces}\ }\textbf {\bibinfo {volume}
  {8}},\ \bibinfo {pages} {18334} (\bibinfo {year} {2016})}\BibitemShut
  {NoStop}%
\end{thebibliography}%

\end{document}